\documentstyle[12pt]{article}
\makeatletter
\def\appendix{\par\clearpage
  \setcounter{section}{0}
  \setcounter{subsection}{0}
  \@addtoreset{equation}{section}
  \def\@sectname{Appendix~}
  \def\theequation{\thesection.\arabic{equation}}
  \def\thesection{\Alph{section}}}
\makeatother
\renewcommand{\theequation}{\thesection.\arabic{equation}}

\begin{document}
\begin{titlepage}
\hskip 11cm \vbox{
\hbox{Budker INP 2000-64}
\hbox{UNICAL-TH 00/6}
\hbox{August 2000}}
\vskip 0.3cm
\centerline{\bf STRONG BOOTSTRAP CONDITIONS$^{~\ast}$}
\vskip 0.8cm
\centerline{V.S.~Fadin$^{a, b~\dagger}$, R.~Fiore$^{c, d~\ddagger}$,
M.I.~Kotsky$^{d~\ddagger}$\footnote{On leave of absence from the Budker Institute
for Nuclear Physics, Novosibirsk, Russia.}, A.~Papa$^{c, d~\ddagger}$}
\vskip .3cm
\centerline{\sl $^a$ Budker Institute for Nuclear Physics, 630090 Novosibirsk,
Russia}
\centerline{\sl $^b$ Novosibirsk State University, 630090 Novosibirsk, Russia}
\centerline{\sl $^c$ Dipartimento di Fisica, Universit\`a della Calabria,}
\centerline{\sl  I-87036 Arcavacata di Rende, Cosenza, Italy}
\centerline{\sl $^d$ Istituto Nazionale di Fisica Nucleare, Gruppo collegato di
Cosenza,}
\centerline{\sl I-87036 Arcavacata di Rende, Cosenza, Italy}
\vskip 1cm
\begin{abstract}
We reformulate the so-called ``strong bootstrap'' conditions for the gluon 
Reg\-gei\-za\-tion in the next-to-leading approximation (NLA), firstly 
suggested by Braun and Vacca, using a different approach, which is not 
based on properties of the eigenstates of the NLA octet BFKL kernel. We write 
the second strong bootstrap condition for the NLA octet impact factors in 
a form which makes clear their dependence on the process. According to this 
condition, the NLA octet impact factors must be given by the product of the 
corresponding Reggeon interaction vertices with a universal coefficient function. 
This function can be used also in the formulation of the first strong 
bootstrap condition for the NLA BFKL kernel in the octet state.
\end{abstract}
\vfill
\hrule
\vskip.3cm
\noindent
$^{\ast}${\it Work supported in part by the Ministero italiano
dell'Universit\`a  e della Ricerca Scientifica e Tecnologica,
in part by INTAS, in part by the Russian Fund of Basic Researches
and in part by the European Research Training Network on QCD and Particle Structure.}
\vfill
$ \begin{array}{ll}
^{\dagger}\mbox{{\it e-mail address:}} &
 \mbox{FADIN ~@INP.NSK.SU}\\
\end{array}
$
\vfill
\vskip -1.5cm
$ \begin{array}{ll}
^{\ddagger}\mbox{{\it e-mail address:}} &
 \mbox{FIORE, KOTSKY, PAPA ~@FIS.UNICAL.IT}\\
\end{array}
$
\vfill
\vskip .1cm
\vfill
\end{titlepage}
\eject

\section{Introduction}
\setcounter{equation}{0}

The gluon Reggeization is one of the remarkable properties of QCD. It plays 
an essential role in the derivation of the BFKL equation~\cite{1}, which
is very important for the description of high energy semi-hard processes in
perturbative QCD. A rigorous proof of the gluon Reggeization has been 
constructed in the leading logarithmic approximation (LLA),
which means summation of the terms $(\alpha_s \ln s)^n$~\cite{4}.
In the next-to-leading approximation (NLA) it has been observed in the first
three orders of perturbation theory~\cite{FFQ}, but its validity to all orders 
of perturbation theory was only assumed in the derivation of the NLA BFKL
equation. It is therefore very important to submit this assumption to careful 
tests.

A very stringent test of the gluon Reggeization is the requirement of its 
compatibility with the $s$-channel unitarity for the NLA elastic processes.
This compatibility has been considered in Ref.~\cite{3}, where two ``bootstrap''
conditions for the NLA octet BFKL kernel and impact factors were obtained. 
Recently, Braun and Vacca~\cite{13,14} suggested a stronger requirement, namely
that the gluon Reggeization should be realized also in the (unphysical) 
particle-Reggeon scattering amplitude with colour octet in the $t$-channel
and negative signature. This requirement leads to ``strong bootstrap''
conditions for the NLA octet BFKL kernel and impact factors.

In this paper we reformulate this strong bootstrap in a different framework. 
Although strictly speaking all the results of this work can be derived from 
those of Refs.~\cite{13,14}, we think that this reformulation can be useful for 
at least two reasons. First, it does not 
rely on properties of the BFKL kernel, such as completeness of the eigenstates,
which are neither evident nor easy to prove. Second, in this approach the
second strong bootstrap condition can be presented in a form which makes
manifest the dependence of the NLA octet impact factors on the particular process.
According to this condition, the NLA octet impact factors must take a very 
simple form, since they can be written as the corresponding Reggeon effective
vertex times a universal coefficient function. This function can be used also 
in the formulation of the first strong bootstrap condition for the NLA BFKL 
kernel in the octet state. We determine this coefficient function and check
the fulfillment of the strong bootstrap conditions related to the quark and gluon
octet impact factors.

The paper is organized as follows: in the next Section we introduce the relevant
quantities and notations and briefly review the bootstrap for the NLA
elastic processes; in Section~3 we consider the strong bootstrap by Braun and 
Vacca and write the related conditions in term of a universal function $R$; 
then in Section~4 we determine this function $R$
and check the strong bootstrap related to the quark and gluon
octet impact factors.

\section{Bootstrap conditions from NLA elastic processes}
\setcounter{equation}{0}

In the BFKL approach~\cite{1} elastic (or quasi-elastic) scattering amplitudes
$\left( {\cal A} \right)^{A^\prime B^\prime}_{AB}$  for the processes $AB
\rightarrow A^\prime B^\prime$ in the Regge kinematics are presented in the
following form~\cite{2}:
\begin{equation}\label{11}
\left( {\cal A} \right)^{A^\prime B^\prime}_{AB} =
\frac{is}{(2\pi)^{D-1}}\sum_{{\cal R},\nu}\langle\Phi^{({\cal R}, \nu)}_{B^\prime
B}|\int^{\delta +
i\infty}_{\delta -i\infty}\frac{d\omega}{\sin(\pi\omega)}\left[ \left(
\frac{-s}{s_0}
\right)^\omega - \tau_{\cal R}\left( \frac{s}{s_0} \right)^\omega \right]
\hat G_\omega^{\cal R}|\Phi^{({\cal R}, \nu)}_{A^\prime A}\rangle\;,
\end{equation}
where $s$ is the squared c.m.s. energy of the colliding particles, which is
assumed to be tending to infinity, the sum is taken over representations 
${\cal R}$ which are contained  in the product of two adjoint 
representations  and over  states $\nu$ from full sets of states 
in these representations,  $\tau_{\cal R}$ is the $t$-channel signature 
which is equal +1 (--1) for symmetric (antisymmetric) representation 
${\cal R}$. The parameter $s_0$ was introduced in Ref.~\cite{2}
to give a convenient definition of the $t$-channel partial wave and is
artificial in the sense that the amplitude does not depend on it in the NLA;
$D = 4 + 2\epsilon$ is the space-time dimension in the dimensional
regularization. Finally, $\Phi$'s are the impact factors of the colliding
particles
and
$G_\omega$ is the Mellin transform of the Green function for the scattering of
two Reggeized gluons,
\begin{equation}\label{12}
\hat G_\omega^{\cal R} = \frac{1}{\omega - \hat{\cal K}^{\cal R}}\; ,
\end{equation}
$\hat{\cal K}$ being the kernel of the BFKL equation. Let us consider the
transverse momentum representation, where ``transverse'' is related to the
plane of initial particle momenta. In this representation, defined by
$$
\hat{\vec q}\: |\vec q_i\rangle = \vec q_i|\vec q_i\rangle\;,\;\;\;
\langle\vec q_1|\vec q_2\rangle =
\vec q_1^{\:2}\vec q_1^{\:\prime \:2}\delta^{(D-2)}(\vec q_1 - \vec q_2)\;,
$$
\begin{equation}\label{13}
\vec q_i^{\:\prime} = \vec q_i - \vec q,\ \ \ \ \ \ \langle A|B\rangle =
\langle A|\vec k\rangle\langle\vec k|B\rangle =
\int\frac{d^{D-2}k}{\vec k^{2}\vec k^{\prime 2}}A(\vec k)B(\vec k)\; ,
\end{equation}
the exact definitions of the impact factors and of the BFKL kernel in the NLA,
\begin{equation}\label{14}
\Phi^{({\cal R}, \nu)}_{A^\prime A}(\vec q_1, \vec q, s_0) = \langle\vec
q_1|\Phi^{({\cal R},\nu)}_{A^\prime A}\rangle\;,
\;\;\;\;\; {\cal K}^{\cal R}(\vec q_2, \vec q_1, \vec q)
= \langle\vec q_2|\hat{\cal K}^{\cal R}|\vec q_1\rangle \; ,
\end{equation}
were given in Ref.~\cite{3}.
In the above equalities $\vec q$ is the transverse momentum transfer for
the amplitude (\ref{11}),  so that we have
\begin{equation}\label{15}
t = q^2 \approx q_\perp^2 = - \vec q^{\:2}.
\end{equation}
Note that throughout this paper $\vec q$ is considered as a fixed parameter
and has nothing in common with the operator $\hat{\vec q}$.
In the derivation of the representation (\ref{11}) a certain Reggeized form
for the production amplitudes was assumed; in particular, for the elastic
amplitude with colour octet representation and negative signature in the
$t$-channel one needs to have
\begin{equation}\label{16}
\left( {\cal A}_{8^-} \right)^{A^\prime B^\prime}_{AB} = \Gamma_{B^\prime
B}^a(s_0)\frac{s}{t}\left[ \left( \frac{s}{s_0} \right)^{\omega(t)} +  
\left( \frac{-s}{s_0} \right)^{\omega(t)} \right]\Gamma_{A^\prime A}^a(s_0),
\end{equation}
where $\Gamma$'s are the particle-Reggeon effective interaction vertices
and $\omega(t)$ is the Reggeized gluon trajectory. The dependence of the
vertices on $s_0$ was indicated explicitly to stress
that it cancels the analogous dependence of the Regge factor in Eq.~(\ref{16})
(the dependence on other possible arguments not being shown). From the
comparison of the term with  ${\cal R} = 8^-$ in Eq.~(\ref{11}) and 
Eq.~(\ref{16}), two
self-consistency relations (the so-called ``bootstrap conditions'') have been
obtained in Ref.~\cite{3}:
\begin{equation}\label{17}
\int\frac{d^{D-2}q_1}{\vec q_1^{\:2}\vec q_1^{\:\prime \:2}}\left[
\int\frac{d^{D-2}q_2}{\vec q_2^{\:2}\vec q_2^{\:\prime \:2}}\langle\vec
q_1|\hat{\cal K}^{(1)}|\vec q_2\rangle  - \omega^{(2)}(t) \right] = 0,
\end{equation}
\begin{equation}\label{18}
-\frac{2ig\sqrt{N}\vec
q^{\:2}}{(2\pi)^{D-1}}\int\frac{d^{D-2}q_1}{\vec q_1^{\:2}
\vec q_1^{\:\prime \:2}}\langle\vec q_1|\Phi^{a(1)}_{A^\prime A}\rangle =
2\omega^{(1)}(t)\Gamma_{A^\prime A}^{a(1)}(s_0) +\omega^{(2)}(t)\Gamma_{A^\prime
A}^{a(0)}.
\end{equation}
In the above relations $\Gamma^{(0)}$ and $\omega^{(1)}$ denote the leading
logarithm approximation (LLA) contributions to the vertex and the gluon
trajectory, respectively. Analogously, $\Gamma^{(1)}$, $\omega^{(2)}$,
${\cal K}^{(1)}$ and $\Phi^{(1)}$ mean the NLA corrections to the
corresponding LLA quantities. We omitted the superscript (or subscript)
${\cal R} = 8^-$, assuming that here and everywhere below all the quantities
belong to this representation.
The conditions~(\ref{17}) and~(\ref{18}) are the bootstrap requirement in the 
NLA in the elastic sector. It is not yet clear whether they are 
sufficient to prove in the NLA the Regge form also of the inelastic 
amplitudes\footnote{In the leading order such form was proved~\cite{4}.},
which is used in the BFKL approach. Nevertheless, these conditions provide a 
strong confirmation of the self-consistency of the NLA BFKL approach. 
In fact, no doubts in the validity of the BFKL approach in the NLA must
remain after the verification that the bootstrap conditions~(\ref{17}) 
and~(\ref{18}) are fulfilled.

The quark part of the first bootstrap condition~(\ref{17}) was analyzed
in~\cite{5} and was found to be satisfied at arbitrary space-time dimension $D$.
Recently, the gluon part of the colour octet kernel was calculated~\cite{6}
in the NLA for $D\rightarrow 4$ and the fulfillment of the bootstrap condition
for this part was proved in this limit~\cite{7}. The calculation of the
kernel at arbitrary $D$ is in progress now~\cite{8}.

The second bootstrap condition (\ref{18}) was checked  for the gluon and quark
impact factors~\cite{9, 10} and was proved to be satisfied at arbitrary space-time
dimension
both for the helicity conserving and non-conserving parts of the impact factors.
Note, that the first bootstrap condition is universal, i.e. process independent, 
and even if it is a quite nontrivial problem to demonstrate its fulfillment,
one has to do it just once. In this respect the second bootstrap condition
looks much worse, because it is process dependent, i.e. any amplitude with
gluon quantum numbers in the $t$-channel has its own bootstrap condition. If one
uses such amplitude in the unitarity relation for a given process, one
should be sure that this bootstrap condition is satisfied.  For example, the NLA
correction to the forward colour singlet BFKL kernel is already 
available~\cite{11, 11a}
and it would be very interesting to apply it for the description of a
real experiment; for this purpose we need to know the forward impact factors
of colourless particles. The
calculation of the virtual photon NLA impact factor  is in progress now~\cite{12}.
Huge mathematical difficulties must be overcome in this calculation, so that
before going on with it, it is desirable to be sure of the self-consistency of
the NLA BFKL approach and, in particular, of the fulfillment of the bootstrap
condition~(\ref{18}) for the impact factor describing the $\gamma \rightarrow
q\bar q$ transition in the photon-Reggeon scattering process, which is involved 
in the calculation. Unfortunately, a direct check of this bootstrap condition is 
a not less complicated problem than the calculation of the photon impact factor 
itself. Therefore one should investigate as much as possible
the bootstrap conditions in general form, without specifying the scattering
process, in order to understand the reasons for the relations~(\ref{17}) 
and~(\ref{18}).

\section{Strong bootstrap conditions}
\setcounter{equation}{0}

Braun and Vacca~\cite{13, 14} suggested stronger bootstrap conditions than those
obtained in~\cite{3}. Their suggestion is based on the requirement of
Reggeization of the unphysical particle-Reggeon scattering amplitude with
colour octet in the $t$-channel and negative signature
\begin{equation}\label{21}
\left( {\cal A} \right)^{A^\prime R^\prime}_{AR} = \frac{i\kappa}{2\pi}\langle\vec
q_1|\int^{\delta + i\infty}
_{\delta - i\infty}\frac{d\omega}{\sin(\pi\omega)}\left[ \left(
\frac{-\kappa}{s_0} \right)^\omega + \left( \frac
{\kappa}{s_0} \right)^\omega \right]\hat G_\omega|\Phi^a_{A^\prime A}\rangle\;,
\end{equation}
where $\kappa$ is the particle-Reggeon squared invariant mass, in the sense that
the amplitude should not contain admixture of eigenstates of the kernel different
from the one corresponding to the gluon trajectory. 
In our reformulation we will not use the language of 
eigenvalues and eigenstates, because, as already discussed, this would require the
justification of some properties of the kernel, such as completeness of the
eigenstates, which are neither evident nor easy to prove. 
In the following we will show briefly how to obtain the strong bootstrap conditions
not assuming these properties.
The meaning of Reggeization was discussed in details in Ref.~\cite{3}
and in the case under consideration it implies
\begin{equation}\label{22}
\left( {\cal A} \right)^{A^\prime R^\prime}_{AR} = {\tilde R}(\vec q_1, \vec q,
s_0)\frac{\kappa}{t}\left[ \left(
\frac{\kappa}{s_0} \right)^{\omega(t)} + \left( \frac{-\kappa}{s_0}
\right)^{\omega(t)} \right]\Gamma
_{A^\prime A}^a(s_0)\;,
\end{equation}
where ${\tilde R}$ can be called the Reggeon scattering vertex. Comparing the 
$\kappa$-channel discontinuities of Eqs.~(\ref{21}) and~(\ref{22}) we get
\begin{equation}\label{23}
\langle\vec q_1|\int^{\delta + i\infty}_{\delta - i\infty}\frac{d\omega}{2\pi i}
\left( \frac{\kappa}{s_0} \right)^\omega\hat G_\omega|\Phi^a_{A^\prime A}\rangle =
\left( \frac{\kappa}{s_0} \right)^{\omega(t)}
\Gamma_{A^\prime A}^a(s_0)\langle\vec q_1|R\rangle\;,
\end{equation}
where
\begin{equation}\label{24}
\langle\vec q_1|R\rangle \equiv R(\vec q_1, \vec q, s_0) \equiv
\frac{\sin(\pi\omega(t))}{-t}{\tilde R}
(\vec q_1, \vec q, s_0)\;.
\end{equation}
Using the relation
\begin{equation}\label{25}
\left( \hat{\cal K}^{(0)} - \omega^{(1)}(t) \right)|\Phi^{a(0)}_{A^\prime
A}\rangle = 0\;,
\end{equation}
which follows from the independence of the Born octet impact factors
$\Phi(\vec q_1, \vec q)$ from $\vec q_1$ and from the form of
the Born octet BFKL kernel (see, for example~\cite{2}), we have in the NLA
$$
\hat G_\omega|\Phi^a_{A^\prime A}\rangle =
\left(\omega - \hat{\cal K}^{(0)}\right)^{-1}|\Phi^{a (1)}_{A^{\prime} A}\rangle
$$
\begin{equation}\label{26}
+\left(\omega^{(1)}(t) - \hat{\cal K}^{(0)}\right)^{-1}
\left( \left(\omega-\omega^{(1)}(t)\right)^{-1}-
\left(\omega - \hat{\cal K}^{(0)}\right)^{-1}\right)
\hat{\cal K}^{(1)}|
\Phi^{a(0)}_{A^\prime A}\rangle\;.
\end{equation}
Therefore we obtain
for the L.H.S. of Eq.~(\ref{23}) with NLA accuracy  (in simplified notations)
$$
\left( \frac{\kappa}{s_0} \right)^{\omega^{(1)}}\langle\vec q_1|\left\{
|\Phi^{(0)}
\rangle + |\Phi^{(1)}\rangle + \ln\left( \frac{\kappa}{s_0} \right)\left[
\hat{\cal K}^{(1)}|\Phi^{(0)}\rangle + \left( \hat{\cal K}^{(0)} -
\omega^{(1)} \right)|\Phi^{(1)}\rangle \right] \right.
$$
\begin{equation}\label{27}
\left. + \sum_{n=0}^{\infty}\frac{\ln^{n+2}\left( \kappa/s_0 \right)}{(n+2)!}
\left( \hat{\cal K}^{(0)} -\omega^{(1)} \right)^{n+1}\left[ \hat{\cal K}^{(1)}|
\Phi^{(0)}\rangle + \left( \hat{\cal K}^{(0)} - \omega^{(1)}\right)|\Phi^{(1)}
\rangle \right] \right\}\; .
\end{equation}
Analogously, the R.H.S. of Eq.~(\ref{23}) takes the form
\begin{equation}\label{28}
\left( \frac{\kappa}{s_0} \right)^{\omega^{(1)}}\langle\vec q_1|
\left\{|R^{(0)}\rangle \left[\Gamma^{(0)}\left(1+\ln\left(
\frac{\kappa}{s_0}\right)\omega^{(2)}\right) +\Gamma^{(1)} \right]
+|R^{(1)}\rangle \Gamma^{(0)} \right\}\;,
\end{equation}
and therefore the compatibility relations are
$$
|\Phi^{(0)}\rangle +|\Phi^{(1)}\rangle =
|R^{(0)}\rangle \left(\Gamma^{(0)} +\Gamma^{(1)}\right)+
|R^{(1)}\rangle\Gamma^{(0)}\;,
$$
$$
\left( \hat{\cal K}^{(1)} - \omega^{(2)} \right)|\Phi^{(0)}\rangle +
\left( \hat{\cal K}^{(0)} - \omega^{(1)}\right)|\Phi^{(1)}\rangle = 0\;,
$$
\begin{equation}\label{29}
\left( \hat{\cal K}^{(0)} - \omega^{(1)} \right)\left[ \hat{\cal K}^{(1)}|
\Phi^{(0)}\rangle + \left( \hat{\cal K}^{(0)}- \omega^{(1)}
\right)|\Phi^{(1)}\rangle
\right] = 0\;.
\end{equation}
We notice now that the third of these relations is not independent: it follows
directly from the second one, taking into account also Eq.~(\ref{25}).
Therefore, restoring the omitted indices of impact factors and vertices, the 
compatibility relations in NLA can be written in the following form
\begin{equation}\label{210}
\left( \hat{\cal K} - \omega(t) \right)|R\rangle = 0\;,
\;\;\;\;\; |\Phi^a_{A^\prime A}\rangle = \Gamma^a_{A^\prime A} |R\rangle\;,
\end{equation}
where $\langle\vec q_1|R\rangle = R(\vec q_1, \vec q, s_0)$ is the universal function defined by the
relations~(\ref{22}) and~(\ref{24}).
Finally, there is one more condition, which is nothing but the generalization
to all orders of the second bootstrap condition~(\ref{18}) for the 
particle-particle scattering amplitudes. Indeed, the amplitude~(\ref{11})
for the case of antisymmetric colour octet state in the $t$-channel
can be represented as the convolution of the impact factor for the particle $B$
and the particle-Reggeon scattering amplitude~(\ref{21}). Using for the latter
amplitude the Regge form~(\ref{22}) and then Eqs.~(\ref{210}), the comparison with
the Regge form~(\ref{16}) leads to the relation
\begin{equation}\label{211}
\frac{\vec q^{\:2}}{(2\pi)^{D-2}}\int\frac{d^{D-2}q_1}{\vec q_1^{\:2}\vec
q_1^{\:\prime \:2}}\langle\vec q_1|
R\rangle^2 = \sin\left( \pi\omega(t) \right)\;.
\end{equation}
In the next section we will see that in the NLA, with account of the second 
from the relations~(\ref{210}), this equation coincides with the 
second bootstrap condition~(\ref{18}).

The set of strong bootstrap conditions, defined by Eqs.~(\ref{210}) and~(\ref{211}) 
is completely equivalent to the one formulated in Refs.~\cite{13, 14}, but 
here it is rewritten in a form which fixes explicitly the process dependence of 
the impact factors: they turn out to be proportional to the corresponding effective 
vertices with the same universal coefficient function $R$. This function
is used also in the formulation of the strong bootstrap condition for the kernel.
To be precise, the Eqs.~(\ref{210}) and~(\ref{211}) can be obtained 
combining the Eq.~(23) of Ref.~\cite{13} and Eqs.~(7) and~(8) of 
Ref.~\cite{14}. It was also stressed in~\cite{13, 14} that the necessity
to assume the Reggeization of unphysical particle-Reggeon scattering amplitudes 
in order to formulate the strong bootstrap conditions can be avoided if one puts 
aside perturbation theory. In our approach this can be done as well, again
without involving the completeness of eigenstates of the kernel. Indeed, 
comparing the $s$-channel discontinuities of Eqs.~(\ref{11}) and~(\ref{16}), 
one easily obtains (in simplified notations)
\begin{equation}\label{212}
\langle\phi_B|\left( \frac{s}{s_0} \right)^{\hat{\cal K} -
\omega}\!|\phi_A\rangle = 1\;,
\end{equation}
where 
\begin{equation}\label{212a} 
|\phi_A\rangle =|\Phi_A\rangle\sqrt{\frac{-t}{(2\pi)^{D-2}
\sin\left( \pi\omega\right)}}\frac{1}{\Gamma_A}\;. 
\end{equation}
An expansion in $\ln\left( s/s_0 \right)$ powers gives
\begin{equation}\label{213}
1 = \langle\phi_B|\phi_A\rangle + \sum_{n=1}^\infty\frac{\ln^n\left( s/s_0
\right)}{n!}\langle\phi_B|\left(
\hat{\cal K} - \omega \right)^n|\phi_A\rangle\;.
\end{equation}
Therefore, for arbitrary $A$ and $B$ (including the case $A=B$) we have
\begin{equation}\label{214}
\langle\phi_B|\phi_A\rangle = 1\;,
\end{equation}
which means that the relation
\begin{equation}\label{215}
|\phi_A\rangle = |\phi\rangle
\end{equation}
is valid for an arbitrary $A$ (since two unit vectors having unity as overlap
coincide).
Now, to remove the higher powers of energy logarithms from Eq.~(\ref{213}), we need
only to put
\begin{equation}\label{216}
\left( \hat{\cal K} - \omega \right)|\phi\rangle = 0\;,
\end{equation}
which gives, together with the previous equality, exactly the strong bootstrap.
Of course, we cannot claim for a non-perturbative consideration, at least
in the BFKL approach. The accurate perturbative analysis of the NLA elastic
bootstrap~\cite{3} shows that in perturbation theory the only necessary
conditions are the soft ones~(\ref{17}) and~(\ref{18}).

\section{Determination of the function $R$}
\setcounter{equation}{0}

To determine the function $R$ introduced in the previous section, we use the
results of Refs.~\cite{10} (see also~\cite{14}) and~\cite{15} for the quark NLA octet impact factor
and the quark-quark-Reggeon effective vertex, correspondingly. For simplicity, we
restrict ourselves to the case of completely massless QCD. Then the quark
impact factor takes the form
$$
\Phi^a_{Q^\prime Q}(\vec q_1, \vec q, s_0) = \Phi^{a(0)}_{Q^\prime Q}\biggl( 1 +
\frac{\omega^{(1)}(t)}{2}
\biggl[ {\tilde K_1} + \left( \left( \frac{\vec q_1^{\:2}}{\vec q^{\:2}}
\right)^\epsilon + \left( \frac{\vec
q_1^{\:\prime \:2}}{\vec q^{\:2}} \right)^\epsilon \right)
\biggl\{ \frac{1}{2\epsilon} + \psi(1+2\epsilon) - \psi
(1+\epsilon)
$$
$$ 
+ \frac{11+7\epsilon}{2(1+2\epsilon)(3+2\epsilon)} -
\frac{n_f}{N}\frac{(1+\epsilon)}{(1+2\epsilon)(3+2\epsilon)}
\biggr\}
$$
\begin{equation}\label{31}
+ \ln\left( \frac{s_0}{\vec q^{\:2}} \right) + 2\psi(1) - 2\psi(1+2\epsilon) -
\frac{3}{2(1+2\epsilon)} - \frac{1}
{N^2}\left( \frac{1}{\epsilon} - \frac{(3-2\epsilon)}{2(1+2\epsilon)} \right)
\biggr] \biggr)\;,
\end{equation}
where $n_f$ is the number of quark flavours and the function ${\tilde K_1}$ has
the following integral representation:
\begin{equation}\label{32}
{\tilde K_1} = \frac{(4\pi)^{2+\epsilon}\Gamma(1+2\epsilon)\epsilon\left( \vec
q^{\:2} \right)^{-\epsilon}}
{4\Gamma(1-\epsilon)\Gamma^2(1+\epsilon)}\int\frac{d^{D-2}k}{(2\pi)^{D-1}}\ln\left
( \frac{\vec q^{\:2}}
{\vec k^{2}} \right)\frac{\vec q^{\:2}}{(\vec k - \vec q_1)^2(\vec k - \vec
q_1^{\:\prime})^2}\;.
\end{equation}
The result of the integration of Eq.~(\ref{32}), in form of expansion in
$\epsilon$,
can be found in Ref.~\cite{10}. The corresponding effective vertex can be presented
as follows:
$$
\Gamma^a_{Q^\prime Q}(s_0) = \Gamma^{a(0)}_{Q^\prime Q}\biggl( 1 +
\frac{\omega^{(1)}(t)}{2}\biggl[ \ln\left(
\frac{s_0}{\vec q^{\:2}} \right) + \frac{1}{\epsilon} + \psi(1-\epsilon) + \psi(1)
- 2\psi(1+\epsilon)
$$
\begin{equation}\label{33}
+ \frac{2+\epsilon}{2(1+2\epsilon)(3+2\epsilon)} - \frac{1}{N^2}\left(
\frac{1}{\epsilon} - \frac{(3-2\epsilon)}
{2(1+2\epsilon)} \right) -
\frac{n_f}{N}\frac{(1+\epsilon)}{(1+2\epsilon)(3+2\epsilon)} \biggr] \biggr)\;.
\end{equation}
Using also the well known expressions for $\Gamma^{(0)}$ and $\Phi^{(0)}$ (see,
for example,~\cite{2}) we obtain  from (\ref{210})
$$
\langle\vec q_1|R\rangle \equiv R(\vec q_1, \vec q) = R^{(0)}\biggl( 1 +
\frac{\omega^{(1)}(t)}{2}\biggl[ {\tilde K_1}
+ \left( \left( \frac{\vec q_1^{\:2}}{\vec q^{\:2}} \right)^\epsilon + \left(
\frac{\vec q_1^{\:\prime \:2}}
{\vec q^{\:2}} \right)^\epsilon - 1 \right)
$$
$$
\times\biggl\{ \frac{1}{2\epsilon} + \psi(1+2\epsilon) - \psi(1+\epsilon) +
\frac{11+7\epsilon}{2(1+2\epsilon)
(3+2\epsilon)} - \frac{n_f}{N}\frac{(1+\epsilon)}{(1+2\epsilon)(3+2\epsilon)}
\biggr\}
$$
\begin{equation}\label{34}
- \frac{1}{2\epsilon} + \psi(1) + \psi(1+\epsilon) - \psi(1-\epsilon) -
\psi(1+2\epsilon) \biggr] \biggr)\;, \;\;\;
R^{(0)} = \frac{-ig\sqrt{N}}{2}\;.
\end{equation}

Let us now discuss the  condition~(\ref{211}). Using the expression~(\ref{34})
for $R^{(0)}$ and the well known integral representation for
$\omega^{(1)}$ (see, for example,~\cite{2}), it can be rewritten in the NLA 
as follows (in simplified notations):
\begin{equation}\label{35}
2\omega^{(1)} + \omega^{(2)} =
4R^{(0)}\int\frac{d^{D-2}q_1}{(2\pi)^{D-1}}\frac{\vec q^{\:2}}{\vec q_1^{\:2}
\vec q_1^{\:\prime \:2}}R\;.
\end{equation}
Then, multiplying both parts with the vertex $\Gamma$, one gets
$$
2\omega^{(1)}\Gamma^{(0)} + 2\omega^{(1)}\Gamma^{(1)} + \omega^{(2)}\Gamma^{(0)} =
4\left( R^{(0)} \right)^2
\Gamma^{(0)}\int\frac{d^{D-2}q_1}{(2\pi)^{D-1}}\frac{\vec q^{\:2}}{\vec
q_1^{\:2}\vec q_1^{\:\prime \:2}}
$$
\begin{equation}\label{36}
+ 4R^{(0)}\int\frac{d^{D-2}q_1}{(2\pi)^{D-1}}\frac{\vec q^{\:2}}{\vec
q_1^{\:2}\vec
q_1^{\:\prime \:2}}\Phi^{(1)}\;.
\end{equation}
The LLA part of the above relation is evidently satisfied and the NLA one gives
exactly the second bootstrap condition~(\ref{18}). It means that this condition is
fulfilled automatically for any process if the conditions~(\ref{210})
and~(\ref{211}) are satisfied, that justifying the term ``strong 
bootstrap conditions''. Let us finally note that it is not necessary to
check the fulfillment of the condition~(\ref{211}) for the function~(\ref{34});
in fact, it was already done when the bootstrap condition~(\ref{18})
for the quark impact factor was checked in Ref.~\cite{10}. Therefore, the above 
discussion completes our formulation of the strong bootstrap conditions. They can 
be presented by the relations~(\ref{210}) and~(\ref{34}). The relation~(\ref{211}) 
is not necessary to include anymore, because the $R$-function~(\ref{34}) does have 
the correct normalization.

We have now the chance to check the second of the strong bootstrap conditions
defined by Eqs.~(\ref{210}) and~(\ref{34}) for the case of gluon scattering,
because the corresponding impact factor and effective vertex are already known
in the NLA and can be found in Refs.~\cite{9} (see also~\cite{14}) 
 and~\cite{16}, correspondingly.
The result for the impact factor is
$$
\frac{\Phi^a_{G^\prime G}(\vec q_1, \vec q, s_0)}{gT^a_{G^\prime G}R^{(0)}} =
\delta_+\biggl\{ 1 + \frac
{\omega^{(1)}(t)}{2}\biggl[ {\tilde K_1} + \left( \left( \frac{\vec
q_1^{\:2}}{\vec
q^{\:2}} \right)^\epsilon +
\left( \frac{\vec q_1^{\:\prime \:2}}{\vec q^{\:2}} \right)^\epsilon - 1
\right)\biggl( \frac{1}{2\epsilon} +
\psi(1+2\epsilon) - \psi(1+\epsilon)
$$
$$
+ \frac{11+7\epsilon}{2(1+2\epsilon)(3+2\epsilon)} -
\frac{n_f}{N}\frac{(1+\epsilon)}{(1+2\epsilon)(3+2\epsilon)}
\biggr) + \ln\left( \frac{s_0}{\vec q^{\:2}} \right) + \frac{3}{2\epsilon} +
2\psi(1) - \psi(1+\epsilon) - \psi(1+2\epsilon)
$$
$$
-\frac{9(1+\epsilon)^2+2}{2(1+\epsilon)(1+2\epsilon)(3+2\epsilon)} 
+ \frac{n_f}{N}\frac{(1+\epsilon)^3+\epsilon^2}{(1+\epsilon)^2(1+2\epsilon)
(3+2\epsilon)} \biggr] \biggr\}  
$$ 
\begin{equation}\label{37} 
- \frac{\delta_-\epsilon\omega^{(1)}(t)} 
{2(1+\epsilon)^2(1+2\epsilon)(3+2\epsilon)}\left( 1 + \epsilon - \frac{n_f}{N} 
\right). 
\end{equation}
Here the notations
\begin{equation}\label{38}
\delta_+ = - \left( {e_\perp^\prime}^*e_\perp \right),\ \ \delta_- = -
{e_\perp^\prime}^*_\mu{e_\perp}_\nu
\left( g^{\mu\nu}_{\perp\perp} - (D-2)\frac{q^\mu_\perp q^\nu_\perp}{q^2_\perp}
\right)
\end{equation}
were used, with $e$ and $e^\prime$ the polarizations of the incoming and
scattered gluons, respectively. Eq.~(\ref{37}), together with the Eq.~(\ref{34}),
gives the expression
$$
\frac{\Phi^a_{G^\prime G}(\vec q_1, \vec q, s_0)}{gT^a_{G^\prime G}R(\vec q_1,
\vec q)} = \delta_+\biggl\{ 1 +
\frac{\omega^{(1)}(t)}{2}\biggl[ \ln\left( \frac{s_0}{\vec q^{\:2}} \right) +
\frac{2}{\epsilon} + \psi(1) +
\psi(1-\epsilon) - 2\psi(1+\epsilon)
$$
$$ 
-\frac{9(1+\epsilon)^2+2}{2(1+\epsilon)(1+2\epsilon)(3+2\epsilon)} 
+ \frac{n_f}{N}\frac{(1+\epsilon)^3+\epsilon^2}{(1+\epsilon)^2(1+2\epsilon)
(3+2\epsilon)} \biggr] \biggr\}  
$$ 
\begin{equation}\label{39} 
- \frac{\delta_-\epsilon\omega^{(1)}(t)} 
{2(1+\epsilon)^2(1+2\epsilon)(3+2\epsilon)}\left( 1 + \epsilon - \frac{n_f}{N} 
\right)
\end{equation}
which must coincide with $\Gamma^a_{G^\prime G}(s_0)/(g T^a_{G^\prime G})$
if the strong bootstrap for impact factors is fulfilled. Simple inspection of
the result~\cite{16} for the vertex confirms its fulfillment.
As for the first strong bootstrap condition (the first of Eqs.~(\ref{210})),
its fulfillment in the part concerning the quark contribution follows from the 
results of Refs.~\cite{13, 17}.
Of course, nothing can be stated as a fact without a
check of the first strong bootstrap condition for the gluon part.

\section{Discussion}
\setcounter{equation}{0}

In this paper we have considered the strong bootstrap conditions, introduced in 
Refs.~\cite{13} and~\cite{14}, which appear as consequences of the hypothesis 
that the unphysical particle-Reggeon scattering amplitudes have also Regge form. 
Our main aims were: first, to show that strong bootstrap conditions do not need 
the assumption of completeness of eigenstates of the octet BFKL kernel;
second, to rewrite them in a form which fixes explicitly their dependence on 
process under consideration.
The strong bootstrap conditions rewritten in such form (see Eqs.~(\ref{210}) 
and ~(\ref{211})) are completely equivalent to the analogous ones of 
Refs.~\cite{13, 14}, as it was explained in the Section~3. We have then 
determined the universal function $R$ using the results for the quark NLA octet 
impact factor~\cite{10} and for the
corresponding effective interaction vertex~\cite{15} and have shown that the
complete set of strong bootstrap conditions can be presented by Eqs.~(\ref{210}) 
and~(\ref{34}) (Eq.~(\ref{211}) is not necessary to include, having 
the function $R$ given by Eq.~(\ref{34}) the proper normalization).
Finally, we have checked the strong bootstrap condition for the gluon NLA impact 
factor and found that it is fulfilled.

The strong bootstrap conditions are satisfied in the LLA~\cite{3} and are likely
to be satisfied also in the NLA, as it may be concluded from the previous section 
and from the results of Refs.~\cite{13, 17}.
This fact is rather intriguing, because the strong bootstrap
is evidently not necessary for the NLA Reggeization of the usual particle
scattering amplitudes, being the soft one (expressed by Eqs.~(\ref{17})
and~(\ref{18})), together with the LLA strong bootstrap, sufficient for it~\cite{3}.
The possibility that the strong bootstrap conditions are necessary for 
the self-consistency of NLA Reggeization in inelastic sector
was mentioned in Ref.~\cite{14}, but, in our opinion, this does not look to be 
the case. The assumption of the NLA Reggeization for the unphysical 
particle-Reggeon scattering amplitude itself cannot be considered a good 
explanation, since it is not yet clear why this requirement should be necessary. 
In any case, there is interesting physics under the strong bootstrap 
(if satisfied) which deserves further clarification.

From a practical point of view, soft bootstrap conditions are now more important
than strong ones. However, since the soft ones are automatically verified if the
strong ones are fulfilled, also the strong bootstrap can be practically used,
especially because it involves one integration less and is therefore easier to 
check. It is also clear that the set of checks of the bootstrap conditions 
for all the possible processes can be replaced by the search of general 
arguments for the NLA Reggeization of the particle-Reggeon scattering amplitudes. 

\vspace{1.0cm}
\underline{Acknowledgment}: Two of us (V.S.F. and M.I.K.) thank the Dipartimento
di Fisica della Universit\`a della Calabria for the warm hospitality while this
work was done. V.S.F. acknowledges the financial support of the Istituto
Nazionale di Fisica Nucleare.

\end{document}